\documentclass{article}
\usepackage{graphicx}
\usepackage{float}
\usepackage{vmargin}

\usepackage{amsmath}
\usepackage{amstext}

\newcommand{\moyen}[1]{\left\langle #1 \right\rangle}
\newcommand{\derivb}[1]{\frac{\partial #1}{\partial \beta}}
\newcommand{\derivmb}[2]{\frac{\partial^{#2} #1}{\partial \beta^{#2}}}
\newcommand{\di}{\mathrm{\; d}}
\DeclareMathOperator*{\Tr}{Tr}

\graphicspath{{diagrams/}{graphics/}}

\newcommand{\diag}[1]{
  \hspace{-1.5mm}
  \begin{tabular}{c}
    \includegraphics[scale=.4]{#1}
  \end{tabular}
  \hspace{-1.5mm}
}

\newcommand{\inlinediag}[1]{
  \hspace{-1.5mm}
  \begin{tabular}{c}
    \includegraphics[scale=.2]{#1}
  \end{tabular}
  \hspace{-1.5mm}
}

\author{Vitor Sessak and R\'emi Monasson \\ Laboratoire de Physique Th\'eorique,  \'Ecole Normale Sup\'erieure\footnote{Unit\'e Mixte du CNRS et de l'\'Ecole Normale Sup\'erieure associ\'ee \`a l'universit\'e Pierre et Marie Curie
Paris 6, UMR 8549. LPTENS-08/53} \\ 24 rue Lhomond -- 75231 Paris Cedex 05 -- France}
\title{Small-correlation expansions for the inverse Ising problem\footnote{To appear in J. Phys. A.}}
\date{September 26, 2008}

\begin{document}

  \maketitle
  
\begin{abstract}
We present a systematic small-correlation expansion to solve the inverse Ising
problem: find a set of couplings and fields corresponding to a given
set of correlations and magnetizations. Couplings are calculated
up to the third order in the correlations for generic magnetizations,
and to the seventh order in the case of zero magnetizations; in
addition we show how to sum some useful
classes of diagrams exactly. The resulting expansion outperforms
existing algorithms on the Sherrington-Kirkpatrick spin-glass model. 
\end{abstract}


\section{Introduction}

Calculating average values of observables given a Hamiltonian is a
general problem in statistical mechanics. This can done either analytically
for a few exactly solvable systems or numerically through simulations
with e.g. Monte Carlo techniques. These techniques give access, 
for not too low temperatures or too big systems,
to the local magnetizations $m_i$ and spin-spin correlations
$c_{ij}$ of an Ising sample, even in the notoriously complex case of spatially
distributed interactions $J_{ij}$ and fields $h_i$ \cite{simu}.
Much less attention has been brought in the physics literature to the
inverse problem, that is, 
calculating the couplings and fields from
the knowledge of the magnetizations and correlations, a problem known
as Boltzmann-machine learning in statistical inference theory
\cite{mckay}. Yet the growing availability of data in
many biological systems of interest as neural assemblies
\cite{bialek1,shlens}, proteins \cite{rama}, gene networks 
\cite{gene}, ... have strengthened the need
for efficient techniques to infer interactions from correlations \cite{mora}.

The purpose of this paper is to present a systematic expansion
procedure to solve the inverse Ising problem. Given a set of
observed magnetizations and correlations we look for the
(a priori non uniform) couplings and fields of the Ising Hamiltonian
reproducing those average observables at equilibrium. Our procedure
is inspired from works by Plefka on mean-field spin glasses
\cite{plefka}, and subsequent results by Georges and Yedidia
\cite{G-Y,livregeorges}, who derived the free-energy of a spin-glass at fixed
magnetization and interactions, performing a Legendre transform of the
free-energy with respect to the fields. Technically speaking our work is an
extension where one more Legendre transform, this time with respect to
the interactions, is carried out to obtain the free-energy at fixed
magnetization and correlations. 

The need for calculating free-energies under some constraints is not new.
One well-known example comes from the
physics of gas or liquids, where one looks for the free-energy of
interacting particles at fixed density and pair correlations
\cite{liquids}. Another example can be found in field theory, where
one is interested in determining the
thermodynamic potential for fixed average
values of the field and two-point correlations \cite{dedom}.
Calculations generally rely on expansions in powers of the
correlations around the non-interacting case which can be exactly
handled. It is important to stress that, in contradistinction with
the above-mentioned examples and most of the existing literature,
our work deals with the case of discrete spin variables and 
non-translationally invariant interactions. 

The plan of the paper is as follows. The general procedure for the
expansion is exposed in Section~\ref{sec_expansion}. Section~\ref{sec_III} is 
devoted to the generic case of non-zero magnetizations while 
Section~\ref{sec_IV} concentrates on the simpler case of zero magnetizations 
where the expansion can be pushed to higher orders. The results for the 
couplings are checked on two standard models: the unidimensional Ising model, 
and the Sherrington-Kirkpatrick (SK) model of a spin-glass. We show that our
procedure for inferring couplings works better than existing methods
for the SK model. The major technicalities are presented in the
Appendices; the reader interested in explicit expressions for the 
couplings given the correlations and magnetizations can skip Section~\ref{sec_expansion}.

\section{Procedure for the Small $c$ Expansion}
\label{sec_expansion}

We consider an Ising model over $N$ spins $\sigma_i=\pm 1$, 
$i=1,\ldots ,N$, with Hamiltonian
\begin{equation}
H(\{S_i\}) = -\sum_{i<j} J_{ij} \sigma_i \sigma_j - \sum_i h_i \sigma_i \ .
\end{equation}
We want to find the values of couplings and the fields,
$J^*_{ij},h^*_i$ such that
the average values of the spins and of the spin-spin correlations
match the prescribed magnetizations $m_i$ and connected correlations 
$c_{ij}$,
\begin{equation}\label{eqmc}
m_i = \frac{\partial  \log Z}{\partial h_i}(\{J^*_{ij}\}, \{h^*_i\})
\quad , \qquad
c_{ij} = \frac{\partial  \log Z}{\partial J_{ij}} 
(\{J^*_{ij}\}, \{h^*_i\}) -m_i\; m_j
\end{equation}
where the partition function (at unit temperature) reads
\begin{equation}\label{partf}
Z(\{J_{ij}\}, \{h_i\}) = \Tr_{\{\sigma_i\}} 
e^{-H(\{\sigma_i\})} \ .
\end{equation}
These couplings and fields are the ones that minimize the entropy of
the Ising model at fixed magnetizations and
correlations\footnote{Note that the minimum may be reached for infinitely large
values of $h_i$ or $J_{ij}$ i.e. as happens for fully
correlated sites $\moyen{\sigma_i \sigma_j}=1$.}, 
\begin{eqnarray}\label{eqG2}
S(\{J_{ij}\},\{h_{i}\}; \{m_i\},\{c_{ij}\} ) &=& 
\log Z(\{J_{ij}\}, \{h_i\}) -
\sum_{i<j} J_{ij} (c_{ij} +m _i m_j)- \sum_i h_i m_i  \\ \nonumber
&=& \log \Tr_{\{\sigma_i\}} \exp\left\{
\sum_{i<j} J_{ij} \left[ (\sigma_i - m_i)(\sigma_j - m_j) -
  c_{ij}\right] + \sum_i \lambda_i (\sigma_i - m_i)\right\}
\end{eqnarray}
where the new fields $\lambda_i$ are simply related to the physical fields
$h_i$ through $\lambda _i= h_i + \sum_j J_{ij} m_j$. 

The calculation of the entropy (\ref{eqG2}) for a given set of
$J_{ij}$ and $\lambda _i$ is, in general, a computationally
challenging task, not to say about its minimization. To obtain a
tractable expression we multiply all (connected) correlations $c_{ij}$ 
in (\ref{eqG2}) by a small parameter $\beta$, which can be
interpreted as a fictitious inverse temperature. 
The calculation of the entropy $S(\{J_{ij}\},\{\lambda_{i}\};
\{m_i\},\{\beta\, c_{ij}\} )$ is straightforward for $\beta =0$ since
spins are uncoupled in this limit. The values of the couplings
and fields minimizing the $\beta=0$ entropy are thus
\begin{equation}
J^*_{ij} (\beta =0) =0 \quad ,\qquad \lambda^*_i(\beta =0) =
h^*_i(\beta =0)  = \tanh^{-1} (m_i) \ .
\end{equation}
Our goal is to expand the couplings and fields in powers of $\beta$;
to each order of the expansion  the couplings and fields will be
functions of the magnetizations and correlations.
Ideally the couplings and fields we are looking for will be obtained when
setting $\beta =1$ in the expansion.

To implement the expansion of $J_{ij}^*$ and $\lambda_i^*$ from
equation~(\ref{eqG2}) we proceed in the following way. First we
define a potential $U$ over the spin configurations at inverse
temperature $\beta$ through
\begin{equation}
U (\{\sigma_i\})= 
\sum_{i<j} J_{ij}^*(\beta) \left[ (\sigma_i - m_i)(\sigma_j - m_j) -
  \beta\; c_{ij}\right]
  + \sum_i \lambda_i^*(\beta) (\sigma_i - m_i) 
  + \sum_{i<j} c_{ij} \int_0^\beta \di \beta' J_{ij}^*(\beta')
  \label{eqU}
\end{equation}
and a modified entropy, compare to (\ref{eqG2}),
\begin{equation}  \label{eeerr}
\tilde S(\{m_i\},\{c_{ij}\}, \beta ) = 
\log \Tr_{\{\sigma_i\}} e^{U(\{\sigma_i\})} \ .
\end{equation}
Notice that $U$ depends on the coupling values $J^*_{ij}(\beta')$
at all inverse temperatures $\beta' < \beta$.
The true entropy (at its minimum) and the modified entropy 
are simply related to each other,
\begin{equation}
S = \tilde S - \sum_{i<j} c_{ij} \int_0^\beta \di \beta' J_{ij}^*(\beta')
\label{realG} \ .
\end{equation}
The modified entropy $\tilde S$ (\ref{eeerr}) has an explicit
dependence on $\beta$ through the potential $U$ (\ref{eqU}), and
an implicit dependence through the couplings and the fields. As the
latter are chosen to minimize $S$ the full derivative of $\tilde S$
with respect to $\beta$ coincides with its partial derivative, and we
get  
\begin{equation}
\label{gconst}
\frac{d\tilde S}{d\beta}= - \sum_{i<j}  c_{ij}\; J_{ij}^*(\beta) + 
\sum_{i<j} c_{ij} \; J_{ij}^* (\beta)=0 \ .
\end{equation}
The above equality is true for any $\beta$.
Consequently $\tilde S$ is constant, and equal to
its $\beta=0$ value, that is, to the entropy of $N$ uncoupled spins with
known magnetizations $\{m_i\}$.

\noindent We now present three facts, shown in the Appendices:
\begin{itemize}
\item[A.] For any integer $k\ge 2$,
\begin{equation} 
\left. \derivmb{\tilde S}{k}\right|_0 =- \sum_{i<j} c_{ij} 
\left. \derivmb{J_{ij}^*}{k-1}\right|_0+ Q_k
\label{anyorder}
\end{equation}
where $Q_k$ is a (known) function of the magnetizations, correlations, 
and of the
derivatives in $\beta=0$ of the couplings $J_{ij}^*$ and fields
$\lambda_i^*$ of order $\le \max (1,k-2)$. See Appendices \ref{ap_A} and \ref{app_mag}.
Recall that $\tilde S$ is constant by
virtue of (\ref{gconst}) thus both sides of (\ref{anyorder}) vanishes. 
\item[B.] For any integer $k\ge 2$
the $k^{th}$ derivative of $\lambda^*_i$ in $\beta=0$ can
be calculated from the magnetizations and the knowledge of the
derivatives in $\beta=0$ of the couplings $J_{ij}^*$ of order
$\le k-1$. See Appendix \ref{derivlambda}. 
\item[C.] The first derivative of the couplings in $\beta=0$ is given by  
\begin{equation}
\label{o1}
\left. \derivb{J_{ij}^*}\right|_0 =\frac{c_{ij}}{(1-m_i^2)(1-m_j^2)} \ .
\end{equation}
See Appendix \ref{app_first_order}.
\end{itemize}

\noindent
Those facts allow us to calculate the derivatives of
the couplings in $\beta=0$ to any order in a recursive way. Let $k\ge
3$. 
From the definition (\ref{eqG2}) of the entropy
\begin{equation}
\frac{\partial S}{\partial c_{ij}}(\{J^*_{ij}\},\{\lambda^*_{i}\}; 
\{m_i\},\{\beta \;c_{ij}\} )= - \beta \; J_{ij}^*  \ .
\end{equation}
Differentiation of the above equation $k$ times with respect to
$\beta$ in $\beta=0$ gives
\begin{equation}
\left. \derivmb{}{k}\right|_0 
\frac{\partial S}{\partial c_{ij}}
(\{J^*_{ij}\},\{\lambda^*_{i}\}; \{m_i\},\{\beta \;c_{ij}\} ) 
= - k\;\left. \derivmb{J_{ij}^*}{k-1}\right|_0 
\ .
\end{equation}
Using relationship (\ref{realG})  we obtain
\begin{equation}
\frac{\partial }{\partial c_{ij}}\left[ \left. \derivmb{\tilde
      S}{k}\right|_0 - \sum_{r<s} c_{rs} \, \left. 
\derivmb{\tilde J_{rs}^*}
{k-1}\right|_0\right]
= - k\;\left. \derivmb{J_{ij}^*}{k-1}\right|_0 
\ .
\end{equation}
We now use that $\tilde S$ is constant and fact A to deduce
\begin{equation}
\left. \derivmb{J_{ij}^*}{k-1} \right|_0 = \frac 1k\; \frac{\partial
  Q_k}{\partial c_{ij}} \ .
\end{equation}
As a consequence the $(k-1)^{th}$ derivative of $J_{ij}^*$ in $\beta=0$ is a
known function of the
derivatives in $\beta=0$ of the couplings $J_{ij}^*$ and fields
$\lambda_i^*$ of order $\le k-2$ (and of the magnetizations and
correlations). Using fact B we express all the derivatives of the fields
in terms of the derivatives of the couplings of order $\le k-2$.
Hence we can compute the $(k-1)^{th}$ derivative of the couplings from
the knowledge of all derivatives with lower orders. The recursive
procedure uses fact C as a starting point to generate all derivatives.

\section{General results for non-zero magnetizations}
\label{sec_III}

\subsection{Explicit expansions of the entropy, couplings and fields}
\label{ss_o_c4}
The procedure exposed in the previous Section has allowed us to
expand the entropy $S$ and the fields $h_i$ up to order $c^4$ and the
couplings up to order $c^3$. Details are given in
Appendix~\ref{details}. We define
\begin{equation}
L_i = 1-m_i^2 \quad ,\qquad  K_{ij} = \frac{c_{ij}}{L_i L_j} \ .
\end{equation}
The entropy reads
\begin{eqnarray}
S &=& 
-\sum_i \left[\frac{1+m_i}{2}\ln\frac{1+m_i}{2} + 
\frac{1-m_i}{2}\ln\frac{1-m_i}{2}\right]
\nonumber\\
&-&
\frac{\beta^2}{2} \sum_{i<j} K_{ij}^2 L_iL_j +
\frac{2}{3} \beta^3 \sum_{i<j}K_{ij}^3 m_i m_j L_i L_j
+ \beta^3 \sum_{i<j<k} K_{ij} K_{jk}K_{ki} L_i L_j L_k\nonumber\\
&-& \frac{\beta^4}{12} \sum_{i<j} K_{ij}^4 \left[1 + 3m_i^2 +3m_j^2 + 9 m_i^2m_j^2\right] L_i L_j
-\frac{\beta^4}{2} \sum_{i<j} \sum_k K_{ik}^2 K_{kj}^2 L_k^2 L_i L_j
\nonumber\\
&-& \beta^4 \sum_{i<j<k<l}
(K_{ij} K_{jk} K_{kl} K_{li} +
  K_{ik} K_{kj} K_{lj} K_{il} + K_{ij} K_{jl} K_{lk} K_{ki})
  L_i L_j L_k L_l
\nonumber\\
&+& O(\beta^5)
\label{eqG}
\end{eqnarray}
The terms in the expansion can be represented diagrammatically. A point
in a diagram represents a spin, and a line represents a $K_{ij}$ link. 
We do not represent the polynomial in the variables $\{m_i\}$ that
multiplies each diagram. Summation over the indices is implicit.
\begin{eqnarray}
S(\{c_{kl}\}, \{m_i\},\beta)  &=&
  -\diag{spin.epsi} -
  \frac{1}{2} \diag{loop1.epsi}
  +\frac{2}{3}\diag{2spins_3.epsi}
  +\diag{loop2.epsi}
  \nonumber \\
  &-&
  \frac{1}{12}\diag{loop_d1.epsi}
  -\frac{1}{2}\diag{3spins_lin.epsi} -
  \diag{loop3.epsi}
\end{eqnarray}
In contradistinction with \cite{G-Y} the
expansion includes non-irreducible diagrams. It should be noted
that, as in \cite{G-Y}, the Feynman rules of these graphs is unknown even
in the $m_i=0$ case, which makes impossible to do the expansion by a simple
enumeration of the diagrams.
The result for $J_{ij}$ is
\begin{eqnarray}
J_{ij}^*(\{c_{kl}\}, \{m_i\},\beta) &=&
\beta K_{ij} -2 \beta^2 m_i m_j K_{ij}^2
- \beta^2 \sum_k K_{jk}K_{ki} L_k\nonumber\\
&+& \frac{1}{3} \beta^3 K_{ij}^3 \left[1 + 3m_i^2 +3m_j^2 + 9 m_i^2m_j^2\right]
+ \beta^3 \sum_{\substack{k \\ (\neq i,\,\neq j)}} K_{ij}
(K_{jk}^2 L_j +K_{ki}^2 L_i)  L_k \nonumber\\
\label{jij}
&+& \beta^3 \sum_{\substack{k, l \\ (k\neq i,\, l\neq j)}} K_{jk} K_{kl} K_{li}L_k L_l
+
 O(\beta^4)
\end{eqnarray}  
We can also represent $J_{ij}^*$ diagrammatically, with the difference that we
connect the $i$ and $j$ sites with a dashed line that do not represent any term
in the expansion:
\begin{eqnarray}
J_{ij}^*  &=&
  \diag{loop1_p.epsi}
  -2 \diag{2spins_3_p.epsi}
  -\diag{loop2_p.epsi}
 \nonumber \\
  &+&
  \frac{1}{3}\diag{loop_d1_p.epsi}
  +\diag{3spins_lin_p1_ij.epsi} +\diag{3spins_lin_p2_ij.epsi} +
  \diag{loop3_p.epsi} \label{diagsJ}
\end{eqnarray}
We end up with the expansion for the `physical' field
\begin{eqnarray}
h_l(\{c_{ij}\}, \{m_i\},\beta) &=&
\frac{1}{2} \ln \left( \frac{1+m_l}{1-m_l}\right)
-\sum_j J_{lj}^* m_j
+ \beta^2 \sum_{j (\neq l)} K_{lj}^2 m_l L_j
\nonumber\\
&-&\frac{2}{3} \beta^3 (1+3m_l^2) \sum_{j (\neq l)} K_{lj}^3 m_j L_j
-2 \beta^3 m_l \sum_{j<k} K_{lj} K_{jk} K_{kl} L_j L_k \nonumber\\
&+& 2 \beta^4 m_l \sum_{i<j} \sum_k K_{ik} K_{kj} K_{jl} K_{li} L_i L_j L_k\nonumber\\
&+&
\beta^4 m_l \sum_{j} K_{lj}^4 L_j \left[1 + m_l^2 + 3 m_j^2 +3m_l^2m_j^2\right]
\nonumber\\
&+& \beta^4 m_l \sum_{i\, (\neq l)} \sum_j K_{ij}^2 K_{jl}^2 L_i L_j^2 
+ O(\beta^5)
\end{eqnarray}
The diagrammatic representation of $h_l$ is very similar to the one of
$S$ (not shown).

\begin{figure}[t]
  \begin{center}
    \includegraphics[width=.6\columnwidth]{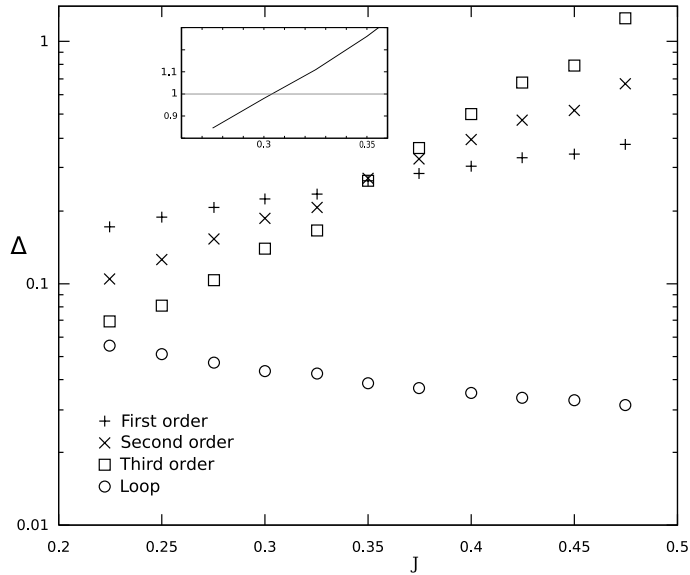}
  \end{center}
\caption{Relative error $\Delta$ (\ref{delta}) on the inferred
  couplings as a function of the inverse temperature $J$ of the
Sherrington-Kirkpatrick model with $N=200$ spins. Monte Carlo
simulations are run over 100 steps, and averages and error bars 
are computed from 100 samples. Top: orders 1,2, 3 of the expansion.
Bottom: expression (\ref{resum1}) which includes the sum over all
loop diagrams. Inset: largest eigenvalue $\Lambda$ of matrix $M$ as a
function of $J$.}
\label{orders}
\end{figure}

We have tested the behaviour of the series on the 
Sherrington-Kirkpatrick model in the paramagnetic phase \cite{SK}. We 
randomly draw a set of $N\times(N-1)/2$ couplings $J_{ij}^{true}$ 
from uncorrelated normal 
distributions of variance $J^2/N$, calculate the correlations and
magnetizations from Monte Carlo simulations, infer the couplings 
$J_{ij}^*$ from the above expansion formulas and compare the outcome to
the true couplings through the estimator
\begin{equation}\label{delta}
\Delta = \sqrt{\frac2{N(N-1)J^2}\sum_{i<j} \left(J_{ij}^*-
    J_{ij}^{true}\right)^2} \ .
\end{equation}
The quality of inference can be seen in Figure 
\ref{orders} for orders (powers of $\beta$)  1,2, and 3.
For large couplings the inference gets
worse when the order of the expansion increases, as could be guessed
from the presence of terms with alternating signs in the expansion,
compare the 2-site loop, triangle, and square in (\ref{jij}), (\ref{diagsJ}).

\subsection{Resummation of loop diagrams}

The divergence coming from the alternate series can be cured by summing all
loop diagrams. A simple inspection shows that each diagram is
multiplied by $\pm 1$ depending on the parity of the number of its links. 
From an algebraic point of view 
\begin{eqnarray}
J_{ij}^{*(\text{loop})} &=&  \beta K_{ij} 
- \beta^2 \sum_k K_{jk}K_{ki} L_k
+ \beta^3 \sum_{k, l} K_{jk} K_{kl} K_{li}L_k L_l + ... \nonumber \\
&=& (L_iL_j)^{-1/2} \left[ (\mathbf{M})_{ij} - (\mathbf{M}^2)_{ij} + 
(\mathbf{M}^3)_{ij} - \cdots \right]
\nonumber \\
&=& (L_iL_j)^{-1/2} \left[\mathbf{M} \cdot (\mathrm{Id} + \mathbf{M})^{-1})\right]_{ij} \label{tap}
\end{eqnarray}
\noindent where $\mathbf{M}$ is the matrix defined by
$\mathbf{M}_{ij} = \beta K_{ij} \sqrt{L_i L_j}$ and
$\mathbf{M}_{ii}=0$. 
Expression~(\ref{tap}) for the coupling 
was already known as a consequence of the TAP equations
(see \cite{TAP1977} and \cite{plefka}), and is exact up to $O(1/N)$ 
corrections  for infinite range models. Our calculation shows how
models with $O(1)$ couplings depart from the TAP expression,
\begin{equation}\label{resum1}
J_{ij}^*(\{c_{kl}\}, \{m_i\}) = J_{ij}^{*(\text{loop})}
 - 2 \beta^2 m_i m_j K_{ij}^2 + 
\frac{2}{3} \beta^3 K_{ij}^3 \left[-1 + 3m_i^2 +3m_j^2 + 3 m_i^2m_j^2\right]
+ O(\beta^4) 
\end{equation}
Figure~\ref{orders} shows how the resummation of loop diagrams 
eliminates the divergence in the relative error $\Delta$ as expected.  
The same phenomenon takes place in the simpler Curie-Weiss model of a
ferromagnet where spins interact through uniform couplings
$J_{ij}=J_0/N$, and the (connected) correlations are of the same order,
$c_{ij}=c/N$. From the relation $N c= \partial m / \partial h$ we can deduce
that the  large-$N$ expression for the coupling
\begin{equation}
J_0 = \frac{c}{1+c} = c - c^2 + c^3 - c^4 + ...
\end{equation}
is an alternating series with radius of convergence $c=1$. This radius
is also given by the condition that the largest eigenvalue of $c_{ij}$
equals 1\footnote{The on-diagonal entries of the correlations are
chosen to be 0 --as is the case for diagonal couplings-- while 
off-diagonal coefficients coincide with
$c_{ij}$.}. This condition applies to the general case too: a necessary
condition for the convergence of equation~(\ref{resum1}) is that the
largest eigenvalue $\Lambda$ of $M$ must be smaller than unity. We plot in the 
Inset of Figure \ref{orders} the behavior of $\Lambda$ as a function of
$J$. It appears that $\Lambda=1$ for $J\simeq .3$, a value comparable
to the intersection point of the lowest order expansions, $J\simeq .35$. 

\begin{figure}[H]
  \begin{center}
      \includegraphics[width=.6\columnwidth]{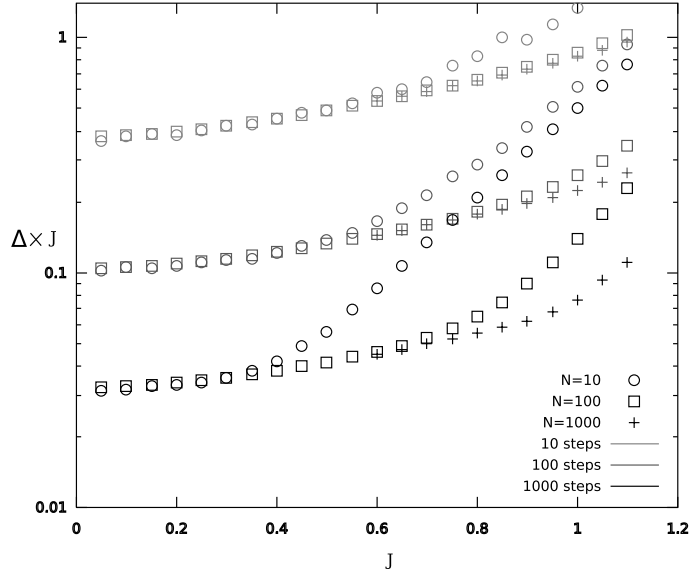}
  \end{center}
\caption{Absolute error $J\times \Delta$ on the inferred couplings 
as a function of the inverse temperature $J$ of the
Sherrington-Kirkpatrick model. Inference is done through formula
(\ref{resum1}), which takes into account all loop diagrams. 
The error decreases with the number of spins and the number of Monte
Carlo steps (shown on the figure).}
      \label{mc2}
  \end{figure}

The apparent large value of the relative error $\Delta$ in Figure
\ref{orders} is not due to the quality of the expansion but to the
the noise in the correlations and magnetizations introduced by the
imperfect sampling of MC simulations. We show in Figure~\ref{mc2} how
the absolute error $J\times\Delta$ decays as the square root of the number
of MC steps, and is roughly independent of $J$ (except close to the
spin-glass temperature $J=1$). As expected, for an infinite
number of MC steps and $N\rightarrow \infty$, the error should vanish.

\subsection{Resummation of two-spin diagrams}
\label{sec_resum2}

Looking carefully at the results of the Section~\ref{ss_o_c4} one can
deduce a general formula for the two spins diagrams,
\begin{eqnarray}
J_{ij}^{*(\text{2-spin})} &=& K_{ij} - 2m_i m_j K_{ij}^2 + 
\frac{1}{3} K_{ij}^3 (1+3m_i^2)(1+3m_j^2) -
4 K_{ij}^4 m_i m_j (1+m_i^2)(1+m_j^2)
\nonumber\\
&+& ... + (-1)^k \frac{1}{k-1} K_{ij}^{k-1} \frac{\moyen{(\sigma_i-m_i)^k}}{1-m_i^2}
\frac{\moyen{(\sigma_j-m_j)^k}}{1-m_j^2}+ ... \nonumber\\
&=&
\frac{1}{4}\ln\left[ 1 + K_{ij}(1+m_i)(1+m_j) \right]
+ \frac{1}{4}\ln\left[ 1 + K_{ij} (1-m_i)(1-m_j) \right] \nonumber\\
&-&\frac{1}{4}\ln\left[ 1 - K_{ij}(1-m_i)(1+m_j) \right] 
-\frac{1}{4}\ln\left[ 1 - K_{ij}(1+m_i)(1-m_j) \right] \label{2spi}
\end{eqnarray}
Where we have used equation (\ref{moyes}) from Appendix~\ref{app_mag} to 
evaluate the averages.
This expression is exact, and was checked by a symbolic
calculation program. Note that in the case of zero magnetization,
(\ref{2spi}) simplifies to $J_{ij}^{*(\text{2 spins})} = \tanh ^{-1} c_{ij}$.

The resummation of all 2-spin diagrams and loop diagrams can be done,
with the result
\begin{equation}\label{duplip}
J_{ij}^{*(\text{2-spin+loop})} = J_{ij}^{*\text{(loop)}}
+ J_{ij}^{*\text{(2-spin)}}- \frac{K_{ij}}{1-K_{ij}^2 L_i L_j}
 \ .
\end{equation}
The last term in (\ref{duplip}) prevents double-counting of 
diagrams of the type $\inlinediag{loop1_p.epsi}$, $\inlinediag{loop_d1_p.epsi}$
(obtained through contraction of $\inlinediag{loop3_p.epsi}$), and is
derived in Appendix \ref{double-counting}. The compact 
expression (\ref{duplip}) contains all the diagrams present 
in (\ref{jij}), in addition to higher order loop 
and 2-spin contributions.

Resummation of all diagrams with a larger number $k$ of spins is
harder. It is done in Section \ref{3spins} in the case of zero
magnetizations and $k=3$. For larger values of $k$ we are not aware of
any closed analytical expression, and resummation can be done by means
of numerical procedures only. An important remark is that
contributions from diagrams with $k$ spins behave as $O(\prod _{i=1}^k
(1-m_i^2))$ when the $m_i$s tend to 1 (or -1) as we show in
Appendix~\ref{app_mag}. This expansion is particularly adapted to the
inference of couplings from strongly magnetized data; a practical application 
can be found in \cite{noi}.

\section{Further results in the zero magnetization case}
\label{sec_IV}

\subsection{Higher order expansions of the entropy, couplings, and
fields}
\label{sec_num}

While the procedure described in Section~\ref{sec_expansion} allows for a
systematic expansion of the couplings in powers of $\beta$ it is
technically involved to do by hand. In this section we find
numerically the expansion up to order $O(\beta^8)$ in the simpler case where
$m_i=0$ for all spins $i$.

We know that the expansion of $S$  up to fifth order is
given by the sum of all diagrams with 5 links or less. More precisely,
\begin{eqnarray}\label{cai}
S^\text{(5th order)} &=& S^\text{(4th order)}
+ a_1 \cdot \diag{o5_1.epsi} 
+a_2 \cdot \diag{o5_2.epsi} +
\nonumber \\
&+&a_3 \cdot \diag{o5_3.epsi}
+a_4 \cdot \diag{o5_4.epsi} 
+a_5 \cdot \diag{o5_5.epsi} 
+ \cdots + \mathrm{O}(c^6)\, .
\end{eqnarray}
As we already know $S^\text{(4th order)}$ from the previous Section
what remains to be found are the $a_i$ coefficients. According to the
procedure outlined in 
Section~\ref{sec_expansion} those coefficients are rational (and in 
particular, for low orders, with a small integer denominator).
Our idea is to find those coefficients from a fit of a numerical solution.

Numerically we minimize the entropy (\ref{eqG2}) for a small number
$N$ of spins 
(not larger than eight). Correlations are arbitrary numbers chosen to
be very small (about $10^{-7}$) since we want the corrections of the
order of $O(c^6)$ to be numerically negligible compared to $O(c^5)$
terms. Of course, when the correlations are very small, so are the
inferred couplings. To estimate the latter to sufficient accuracy 
we have performed our calculations with a unusual
large number of decimal units ($\approx 400$).
A computer program, at each step $l$, randomly chooses 
the couplings $c_{ij}^l$ and numerically  evaluates  the 
corresponding entropy  $S_l$ and correlations $c_{ij}^l$ through an exact
enumeration over the $2^N$ spin configurations. Then it calculates
\begin{equation}
  D = \sum_{l=1}^L 
  \left[S_l - S^\text{(5th order)}(c_{ij}^l)\right]^2
\end{equation}
over a large number $L$ of random samples.
This quantity is quadratic in the coefficients $a_i$, so its minimum
can be easily obtained, and we could deduce that the coefficients 
in the expansion (\ref{cai}) are all zero. Using this procedure, order by 
order, we have determined the 
following expansion for $J_{ij}^*$  (where the coefficients found 
numerically differed from the rational
fractions listed below by less than $10^{-10}$):
\begin{eqnarray}
J_{ij}^*  &=& J_{ij}^{*\text{(2-spin+loop)}} + \diag{loop_d2_p.epsi}
  - \frac{4}{3}\diag{o7_p1.epsi}
  -4 \left(\diag{o7_p2_ij.epsi} + \diag{o7_p3_ij.epsi}\right)\nonumber\\
  &+& 2 \left(\diag{3spins_lin2_p1_ij.epsi}+\diag{3spins_lin2_p2_ij.epsi}\right)
  + 16 \diag{o8_3_p1.epsi}
  + 8\left(\diag{o8_3_p2_ij.epsi}+\diag{o8_3_p3_ij.epsi}\right)\nonumber\\
  &-& 2 \diag{loop_d3_p.epsi}
  -4 \diag{o8_2_p1.epsi}
  -4 \diag{o8_2_p2.epsi} + O(c^8)
  \label{exp_numer}
\end{eqnarray}

We can note the absence of any term with five or more spins in this expansion.
We suspect that the lowest order diagram in this expansion for a given number 
spins is the loop with double links. In particular, the first diagram with five
spins would be $\inlinediag{loop_d5.epsi}$, which is not present in 
(\ref{exp_numer}) since it is $O(c^9)$.

\subsection{Three spins summation for zero magnetization}
\label{3spins}
For three spins and zero magnetizations the entropy (\ref{eqG2}) can
be minimized exactly with a symbolic algebra program, with the
following results for the couplings
\begin{eqnarray}
J_{ij}^{* \text{(3-spin)}} = \frac{1}{4} \sum_{k\, 
(\neq i, j)} &&\left\{ \log\left[\frac{1 + c_{ij} - c_{ik} -
    c_{jk}}{1 - c_{ij} - c_{ik} + c_{jk}} \right]\right. \nonumber \\
     &-&  \left. \log \left[\frac{1 - c_{ij} + c_{ik} - c_{jk}}{1 -
     c_{ij} - 
     c_{ik} + c_{jk}}\right] + 
         \log \left[\frac{1 + c_{ij} + c_{ik} + c_{jk}}{1 - c_{ij} -
	 c_{ik} + c_{jk}}\right] \right\}
\end{eqnarray}
Following the same lines as in Section \ref{sec_resum2} we gather our
previous results in the $m_i=0$ case under the
form\footnote{Calculations to avoid double-counting are similar to the
ones shown in Appendix \ref{double-counting}, with a $3\times3$
instead of $2\times 2$ matrix, and are not shown.},
\begin{eqnarray}\label{all}
	 J_{ij}^{*\text{(2-spin+loop+3-spin)}} &=& 
 J_{ij}^{*\text{(2-spin+loop)}} + J_{ij}^{* \text{(3-spin)}} \nonumber \\
		       &-& \sum_{k\, (\neq i, j)} \left\{
J_{ij}^{*\text{(2 spins)}}+\frac{c_{ij} - c_{ik} c_{jk}}{
   1-c_{ij}^2-c_{ik}^2-c_{jk}^2 +
			       2 c_{ij}c_{jk}c_{ki}} -
\frac{c_{ij}}{1-c_{ij}^2}
			       \right\}\, .
\end{eqnarray}

\subsection{Check on the one-dimensional Ising model}
\label{1dising}
We consider a unidimensional Ising model with uniform coupling $J$
between nearest neighbours\footnote{The case of non-uniform couplings
  varying from link to link can be treated along the same lines.}. We have
\begin{equation}
c_{ij} = \left|\tanh J\right|^{|i-j|}
\end{equation}
\begin{equation}
J_{ij}^{*\text{(2-spin)}} = \tanh^{-1}\bigg(
\left|\tanh J\right|^{|i-j|}\bigg)
\end{equation}
\noindent We can see from the above formula that the sum of all 2-spin diagrams
infer the correct value of the couplings $J_{i,i+1}$, but give a non-zero value
for the other ones. We can also evaluate the sum of loop
diagrams (with $c=\tanh J$):
\begin{equation}
J_{ij}^{*\text{(loop)}} = 
\frac{c}{1-c^2} (\delta_{i,i+1} + \delta_{i,i-1})  \, ,
\end{equation}
where $\delta_{i,j}$ is the Kronecker function.
We obtain a zero contribution for non-neighbouring sites, but
an erroneous values for the nearest-neighbour coupling
$J_{i,i+1}$. If we consider the contributions from both 2-spin and loop
diagrams,
\begin{eqnarray}
J_{ij}^\text{(2-spin+loop)} &=& 
 J (\delta_{i,i+1} + \delta_{i,i-1}) +
\left[\tanh c_{ij} - \frac{c_{ij}}{1 - c_{ij}^2}\right](1-\delta_{i,i+1})
(1-\delta_{i,i-1}) \nonumber \\
&=& J (\delta_{i,i+1} + \delta_{i,i-1}) + O(c^6) \ , 
\end{eqnarray}
which is correct to the order $c^6$. The next contribution to the
couplings coming from the
expansion~(\ref{exp_numer}) corresponds to $\inlinediag{loop_d2_p.epsi}$, whose
leading term is indeed proportional to
$c_{i,i+2} \cdot c_{i,i+1}^2  \cdot c_{i+1,i+2}^2 = c^6$.

We may also want to understand how $J_{i,i+2}$ converges to zero. 
Using geometric series calculations one can evaluate
\begin{eqnarray}
\diag{alphabeta.epsi} = c^{2 \gamma + \alpha+\beta} \cdot \frac{1+ c^{2\alpha} + c^{2\beta}}{1 - c^{\alpha+\beta}}\\
\diag{loop_d3_p.epsi} = 2 c^{10} \frac{1+c^4}{1-c^4} + c^{14} \frac{2+c^8}{(1-c^4)^2}
\end{eqnarray}
Performing the whole summation in (\ref{exp_numer}) we find that
$J_{i,i+2}=O(c^8)$, which is consistent
with the first missing contribution from the expansion,
$\inlinediag{o9_triang.epsi}$.

\subsection{Application to the Sherrington-Kirkpatrick model}
\label{skexact}

We have seen above that the error on the inferred couplings for the
Sherrington-Kirpatrick model is essentially due to the noise
in the MC estimates of the correlations and magnetizations.
To avoid this source of noise we now evaluate the error due to our
truncated expansion using a program that calculates $c_{ij}$ through
an exact enumeration of all $2^N$ spin configurations. We are limited
to small values of $N$ (10, 15 and 20). However the case of a small 
number of spins is particularly interesting because, for the SK model, 
the summation of loop diagrams is exact in the
limit $N \rightarrow \infty$. The importance of terms in our
expansions not included in the loop resummation is thus better 
studied at small N.

\begin{figure}[H]
 \begin{center}
   \includegraphics[width=.6\columnwidth]{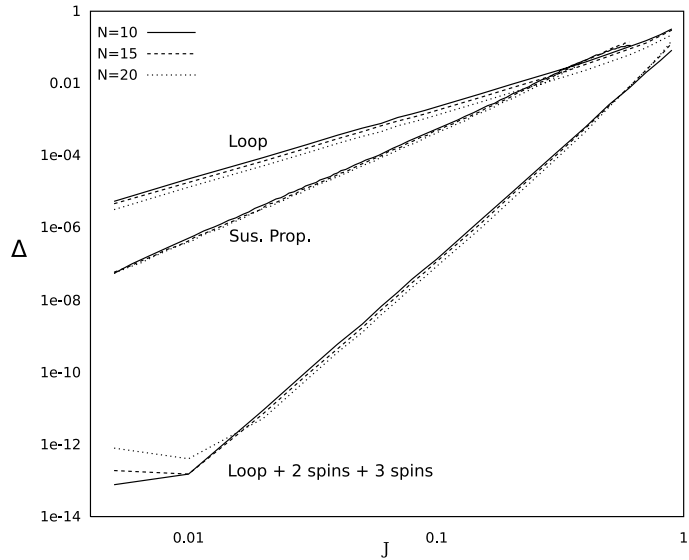}
 \end{center}
\caption{Relative error $\Delta$ (\ref{delta}) as a function of $J$ for the SK
  model for our resummation $J_{ij}^\text{(2 spin + loop + 3 spin)}$ 
(\ref{all}) compared to the Susceptibility Propagation method of M\'ezard and 
Mora \cite{mora} and loop resummation $J_{ij}^\text{(loop)}$ 
(\ref{resum1}).}
 \label{orders2}
\end{figure}

Results are shown in Figure \ref{orders2}. The error is remarkably
small for weak couplings, and get dominated by finite-digit accuracy 
($10^{-13}$) in this limit. Not surprisingly it behaves better than
simple loop resummation, and also outperforms the
message-passing-based method recently introduced in \cite{mora}.

\section{Perspectives}

As we saw in Sections~\ref{1dising} and~\ref{skexact} the expansion 
method introduced in
this paper works well for both the Sherrington-Kirkpatrick
sping-glass --an infinite dimensional system,
with very dilute couplings-- and the unidimensional Ising --with only a few but
strong couplings per site-- models. It would be interesting to investigate how
accurate our method is for `Small-World'-like interaction
networks, which have both kinds of couplings \cite{sworld}.

In principle the assumption of binary-valued spins ($\sigma_i=\pm 1$) is not
central to our expansion and could be straightforwardly released to tackle the
case of Potts models, where each spin can be in $q$ possible states
($\sigma_i = 1,...,q$). Such a generalization
would make the method useful to connect with biological problems 
involving amino-acids \cite{rama}.

Finally our expansion breaks down with the onset of the spin-glass
phase as can be seen from Figure \ref{orders2}. The failure of our
method (and of other existing algorithms) is not
surprising. Correlations and magnetizations have a physical meaning
when there is a single pure state. In presence of more than one phases
Gibbs averages have indirect significance. A well-known example is the
ferromagnet at low temperature and zero field where two equally-likely
phases of opposite magnetizations $\pm m_s$ exist, and the resulting
Gibbs magnetization $m$ truly vanishes (for any finite
$N$). Work is in progress to extend our expansion technique to this
multiple-phase regime.

\vskip 1cm
{\bf Acknowledgments:} We are grateful to S. Cocco and S. Leibler for 
very useful and insightful discussions. We also thank S. C. for the critical
reading of the manuscript. This work was
partially funded by the Agence Nationale de la Recherche JCJC06-144328 
and the Galileo 17460QJ contracts.
\appendix

\section{Details of the small-$\beta$ expansion}
\label{ap_A}
Let $O$ be an observable of the spin configuration (which can
explicitly depend on the inverse temperature $\beta$), and 
\begin{equation}
\moyen{O} = \frac 1Z \Tr_{\{ \sigma_i\}} O \;e^U
\end{equation} 
its average value, where $U$ is defined in (\ref{eqU}), and
$Z=\exp(\tilde S)$.
The derivative of the average value of $O$ fulfills the following
identity,
\begin{equation}
  \label{eqderiv}
\derivb{\moyen{O}} = \frac{1}{Z} 
\Tr_{\{\sigma_i\}}  \left[\derivb{O}+O\derivb{U} \right] e^U - 
\frac{1}{Z^2} \derivb{Z} \Tr_{\{\sigma_i\}}  Oe^U = \moyen{\derivb{O}} + \moyen{O\derivb{U}}
\end{equation}
where the term in $Z^{-2}$ vanishes as a consequence of (\ref{gconst}).

\label{details}
\subsection{First order expansion}
\label{app_first_order}
Using (\ref{eqderiv}) and (\ref{gconst}),
\begin{equation}
0 = \derivmb{\tilde{S}}{2}  = \derivb{}\moyen{\derivb{U}} = 
\moyen{\derivmb{U}{2}} + \moyen{\left(\derivb{U}\right)^2} \label{g2}
\end{equation}
and by using the explicit form of $U$ given in (\ref{eqU}):
\begin{equation}
\left. \derivmb{\tilde{S}}{2} \right|_0 = -\sum_{i<j} c_{ij} \left. \derivb{J_{ij}^*}\right|_0 + \sum_{i<j} \left(\left. \derivb{J_{ij}^*}\right|_0\right)^2 (1-m_i^2)(1-m_j^2) + \sum_i \left(\left. \derivb{\lambda_i^*} \right|_0\right)^2 (1-m_i^2)
\end{equation}

In Appendix~\ref{derivlambda} we show that
$\left. \derivb{\lambda_i^*} \right|_0 =0$. As $\tilde S$ is constant
we end up with the
following algebraic equation for the first order derivative of
$J_{ij}^*(\beta)$,
\begin{equation}
0=  -\sum_{i<j} c_{ij} \left. \derivb{J_{ij}^*}\right|_0 + \sum_{i<j}
 \left(\left. \derivb{J_{ij}^*}\right|_0\right)^2 (1-m_i^2)(1-m_j^2)
 \ .
\end{equation}
The only non-zero solution of the above equation, symmetric under
index permutations, is the announced result (\ref{o1}).

\subsection{Second order expansion}

Using (\ref{eqderiv}) and~(\ref{g2})
\begin{equation}
  \label{g3}
0= \derivmb{\tilde{S}}{3} = \derivb{}\left[\moyen{\derivmb{U}{2}} + 
\moyen{\left(\derivb{U}\right)^2}\right] = 
\moyen{\derivmb{U}{3}} + 3\moyen{\derivmb{U}{2}\derivb{U}} + \moyen{\left(\derivb{U}\right)^3}
\end{equation}

A straightforward calculation gives (where we omit for clarity the
notation $|_0$ and the ${}^*$ subscript from $J_{ij}$ and $\lambda_i$)

\begin{eqnarray}
\moyen{\derivmb{U}{3}}_0 &=& - 2\sum_{i<j} \derivmb{J_{ij}}{2} c_{ij} \\
\moyen{\derivmb{U}{2}  \derivb{U}}_0 &=&
\sum_{i<j} \derivmb{J_{ij}}{2} \derivb{J_{ij}} L_i L_j +
\sum_{i} \derivmb{\lambda_i}{2} \derivb{\lambda_i} L_i \\
\moyen{\left( \derivb{U}  \right)^3}_0 &=& 
6 \sum_{i<j<k} \derivb{J_{ij}} \derivb{J_{jk}} \derivb{J_{ki}} L_i
L_j L_k +
\nonumber \\ &+&
\sum_{i<j} \left(\derivb{J_{ij}}\right)^3 4 m_i m_j L_i L_j
+6 \sum_{i<j} \derivb{J_{ij}} \derivb{\lambda_i}
\derivb{\lambda_j}L_i L_j
\end{eqnarray}  

Using (\ref{g3}), results from Appendix~\ref{derivlambda} for the
expressions of the derivatives of $\lambda_i$ in $\beta=0$, and
(\ref{o1}) we obtain (\ref{anyorder}) for $k=3$  with 
\begin{equation}
Q_2 = -
4 \sum_{i<j} \frac{c_{ij}^3 m_i m_j}{(1-m_i^2)^2(1-m_j^2)^2} -
6 \sum_{i<j<k} \frac{c_{ij} c_{jk} c_{ki}}{(1-m_i^2)(1-m_j^2)(1-m_k^2)}
\end{equation}
from which we deduce
\begin{equation}
\left. \derivmb{S}{3} \right|_0 = 
4 \sum_{i<j} K_{ij}^3 m_i m_j L_i L_j +
6 \sum_{i<j<k} K_{ij} K_{jk} K_{ki} L_i L_j L_k
\end{equation}
and
\begin{equation}
\left. \derivmb{J_{ij}}{2}\right|_0
=
-4 m_i m_j K_{ij}^2 - 2 \sum_{k (\neq i,\, \neq j)} K_{jk} K_{ki} L_k
\ .
\end{equation}

\subsection{Third order expansion}

The procedure to derive the third order expansion for the coupling is
identical to the second order one. We start from
\begin{equation}
0 = \derivmb{\tilde{S}}{4} = \moyen{\derivmb{U}{4}} + 
3 \moyen{\left(\derivmb{U}{2}\right)^2} + 4 \moyen{\derivmb{U}{3}\derivb{U}}
+ 6\moyen{\left(\derivb{U}\right)^2\derivmb{U}{2}} +
\moyen{\left(\derivb{U}\right)^4}
\end{equation}
and evaluate each term in  the sum:
\begin{eqnarray}
\moyen{\derivmb{U}{4}}_0 &=&
-3 \sum_{i<j} \left. \derivmb{J_{ij}}{3}\right|_0 K_{ij} L_i L_j
\\
\moyen{\left(\derivmb{U}{2}\right)^2}_0 &=&
\sum_{i<j} \left(\derivmb{J_{ij}}{2}\right)^2 L_i L_j +
\sum_i \left(\derivmb{\lambda_i}{2}\right)^2 L_i + 
\left[ \sum_{i<j} K_{ij}^2 L_i L_j \right]^2
\\
\moyen{\derivmb{U}{3}\derivb{U}}_0 &=& 
\sum_{i<j} K_{ij} \derivmb{J_{ij}}{3} L_i L_j
\\
\moyen{\left(\derivb{U}\right)^2\derivmb{U}{2}}_0 &=&
2\sum_{i<k} \sum_j K_{ij}K_{jk}\derivmb{J_{ki}}{2} L_i L_j L_k
+ 4\sum_{i<j} K_{ij}^2 \derivmb{J_{ij}}{2} m_i m_jL_i L_j
\nonumber\\
&+& \sum_i \sum_j K_{ij}^2 \derivmb{\lambda_i}{2}(-2m_i)L_i L_j
- \moyen{\left(\derivb{U}\right)^2}_0 \sum_{i<j} K_{ij}^2 L_i L_j
\\
\moyen{\left(\derivb{U}\right)^4}_0
&=& \sum_{i<j} K_{ij}^4 (3m_i^2+1)L_i (3m_j^2+1)L_j 
+ 3 \sum_{i<j, \,k<l\, (k\neq i,\, l\neq j)}
K_{ij}^2 K_{kl}^2 L_i L_j L_k L_l +
\nonumber\\
&+& 6 \sum_{i<k}\sum_j K_{ij}^2 K_{jk}^2 (3m_j^2+1) L_i L_j L_k +
\nonumber\\
&+& 12 \sum_{i<j<k} K_{ij} K_{jk} K_{ki} L_i L_j L_k \left[
4m_i m_j K_{ij} + 4m_im_k K_{ik} + 4m_km_i K_{ki}\right] +
\nonumber\\
&+& 3\sum_{i,j,k,l \, (\neq)}
K_{ij} K_{jk} K_{kl} K_{li} 
L_i L_j L_k L_l
\end{eqnarray}
Using the results from Appendix B we can write all the terms above in the
same form
\begin{eqnarray}
-3\left[ \sum_{i<j} K_{ij}^2 L_i L_j \right]^2 &=&
- 3 \sum_{i<j,\,k<l\, (k\neq i,\, l\neq j)} K_{ij}^2 K_{kl}^2 L_i L_j L_k L_l
\nonumber\\
&-&
6 \sum_{i<j} \sum_k K_{ik}^2 K_{kj}^2 L_i L_j L_k^2
- 3 \sum_{i<j} K_{ij}^4 L_i^2 L_j^2
\\
12\sum_{i<j}\sum_k K_{ik}K_{kj}\derivmb{J_{ij}}{2} L_i L_j L_k &=&
-48 \sum_{i<j}  \sum_k K_{ij}^2 K_{ik} K_{kj} m_i m_j L_i L_j L_k -
\nonumber\\
&-& 12 \sum_{i,j,k,l \, (\neq)}
K_{ij} K_{jk} K_{kl} K_{li}L_i L_j L_k L_l
\nonumber\\
&-& 24 \sum_{i<j} \sum_k K_{ik}^2 K_{kj}^2 L_i L_j L_k^2
\\
\sum_{i<j} K_{ij}^2\derivmb{J_{ij}}{2} m_i m_j L_i L_j &=&
 -4 \sum_{i<j} K_{ij}^4 m_i^2 m_j^2 L_i L_j 
- 2\sum_{i<j} \sum_k K_{ij}^2 K_{jk} K_{ki} m_i m_j L_iL_jL_k
\\
3\sum_{i<j} \left(\derivmb{J_{ij}}{2}\right)^2 L_i L_j &=&
48 \sum_{i<j} K_{ij}^4m_i^2 m_j^2L_i L_j +
48 \sum_{i<j} \sum_k K_{ij}^2 K_{ik} K_{kj} m_i m_j L_i L_j L_k +
\nonumber\\
&+& 6 \sum_{i,j,k,l \, (\neq)}
K_{ij} K_{jk} K_{kl} K_{li} L_iL_jL_kL_l
\nonumber\\
&+& 12 \sum_{i<j} \sum_k K_{ik}^2 K_{kj}^2 L_k^2 L_i L_j
\\
6\sum_i \sum_j K_{ij}^2 \derivmb{\lambda_i}{2} (-2m_i)L_iL_j &=&
-24 \sum_i \sum_j K_{ij}^4 m_i^2 (1-m_j^2) L_iL_j
\nonumber \\
&-& 48 \sum_{i<j} \sum_k K_{ik}^2 K_{kj}^2 m_k^2 L_iL_jL_k
\\
3 \sum_k \left(\derivmb{\lambda_k}{2}\right)^2 L_k &=&
24 \sum_{i<j} \sum_k K_{ik}^2 K_{kj}^2 m_k^2 L_iL_jL_k
+ 12 \sum_i \sum_j K_{ij}^4 m_i^2 L_i L_j^2 
\end{eqnarray}
Again we find equation (\ref{anyorder}) with 
\begin{eqnarray}
Q_3 &=& -\sum_{i<j} K_{ij}^4 \left[(3m_i^2+1) (3m_j^2+1) 
-48 m_i^2m_j^2\right]L_iL_j
\nonumber \\
&+& 12 \sum_{i<j}\sum_k K_{ik}^2 K_{jk}^2 L_i L_j L_k^2
+ 3 \sum_{i,j,k,l\,(\neq)}
K_{ij} K_{jk} K_{kl} K_{li}L_i L_j L_k L_l
\nonumber \\
&+& 12 \sum_i \sum_j K_{ij}^4 m_i^2  L_iL_j^2
+ 3 \sum_{i<j} K_{ij}^4 L_i^2 L_j^2
\end{eqnarray}
which gives the fourth order contribution to the entropy,
\begin{eqnarray}
\derivmb{S}{4}  &=& -2\sum_{i<j} K_{ij}^4 \left[1 + 3m_i^2 +3m_j^2 + 9 m_i^2m_j^2\right] L_i L_j
-12 \sum_{i<j} \sum_k K_{ik}^2 K_{kj}^2 L_k^2 L_i L_j
\nonumber\\
&-& 24 \sum_{i<j<k<l}
(K_{ij} K_{jk} K_{kl} K_{li} +
  K_{ik} K_{kj} K_{lj} K_{il} + K_{ij} K_{jl} K_{lk} K_{ki})
  L_i L_j L_k L_l
\end{eqnarray}
and the third order contribution to the coupling,
\begin{eqnarray}
\left. \derivmb{J_{ij}}{3}\right|_0 &=& 
2 K_{ij}^3 \left[1 + 3m_i^2 +3m_j^2 + 9 m_i^2m_j^2\right]
+6 \sum_{k\, (\neq i,\,\neq j)} K_{ij}
(K_{jk}^2 L_j +K_{ki}^2 L_i)  L_k
+
\nonumber\\
&+& 6 \sum_{\substack{k, l \\ (k \neq i, l\neq j)}} K_{jk} K_{kl} K_{li} L_k L_l \, .
\end{eqnarray}

\section{Derivatives of $\lambda^*_i$ in $\beta=0$}
\label{derivlambda}

Since $m_i$ and $h_i$ are conjugated thermodynamic variables it is 
natural to evaluate
\begin{eqnarray}
\frac{\partial \tilde{S}}{\partial m_k} &=& \moyen{\frac{\partial U}{\partial m_k}}
= \sum_{i<j} c_{ij} \int_0^\beta \di \beta' 
\frac{\partial J_{ij}^*(\beta')}{\partial m_k}\nonumber \\
&-& \sum_{i<j} J_{ij}^*(\beta)\moyen{(\sigma_i - m_i) \delta_{jk} +
(\sigma_j - m_j)\delta_{ik}} + \sum_{i<j} \frac{\partial J_{ij}^*}{\partial m_k}
\moyen{ (\sigma_i-m_i)(\sigma_j-m_j)} - \lambda_k^*(\beta) \nonumber \\
&=& -\lambda_k^*(\beta)+ \sum_{i<j} c_{ij} \int_0^\beta \di \beta' 
\frac{\partial J_{ij}^*(\beta')}{\partial m_k}
\end{eqnarray}
As the modified entropy is independent of $\beta$,
\begin{equation}
\frac{\partial \tilde{S}}{\partial m_k} =
\left. \frac{\partial \tilde{S}}{\partial m_k} \right|_0
= -\lambda_k^*(0) ={\tanh}^{-1} (m_k) = 
\frac{1}{2} \ln\left(\frac{1+m_k}{1-m_k}\right)
\end{equation}
where we used the well-known result for the entropy of
uncorrelated spins. We can then deduce the formula, valid for any $\beta$:
\begin{equation}
\label{eq43}
\lambda_k^*(\beta) = \frac{1}{2} \ln\left(\frac{1-m_k}{1+m_k}\right)
+ \sum_{i<j} c_{ij} \int_0^\beta \di \beta' 
\frac{\partial J_{ij}^*(\beta')}{\partial m_k}\ .
\end{equation}
It is now straightforward to deduce the expansion of $\lambda_i^*$ to the
order $O(\beta^k)$ from the expansion of $J_{ij}^*$ to the order
$O(\beta^{k-1})$. In particular,
\begin{equation}
\left. \derivb{\lambda_k^*}\right|_0 = 0
\end{equation}
and using the order $O(\beta)$ of the $J_{ij}^*$ expansion,
\begin{equation}
\left. \derivmb{\lambda_k^*}{2}\right|_0 = 2 m_k \sum_{i} K_{ik}^2 L_i \, .
\end{equation}

\section{Large magnetization expansion}
\label{app_mag}
Equation (\ref{jij}) suggests that to expand $J_{ij}^*$ to the order
of $(L_i)^k$ one has to sum all the diagrams with up to $k+2$ spins. 
This statement is true if the  expansion for $J_{ij}^*$ is of the form
\begin{equation}
\label{forma}
J_{ij}^* = A_{ij} + \sum_k L_k A_{ijk} + \sum_k \sum_l L_k L_l A_{ijkl} + ...
\end{equation}
where the coefficients $A_{i_1 i_2 ... i_n}$ are polynomials in the
couplings $K_{i_\alpha i_\beta}$ and the magnetizations
$m_\alpha$ ($\alpha,\beta < n$).
In the following we will show that the above statement is true to 
any order of the expansion in $\beta$ by recurrence. 
First of all, from (\ref{eq43}) we see that if $J_{ij}^*$ is of
the form~(\ref{forma}) up to the order $k$, so is $\lambda_i^*$ to the same
order.

As we saw in section~\ref{sec_expansion}, to find an equation
 for $\derivmb{S}{k}$, one must evaluate $  \derivmb{\tilde{S}}{k+1}$. Using
Eq.~\ref{eqderiv}, we can write

\begin{equation}
\label{ss}
  \derivmb{\tilde{S}}{k+1} = \moyen{\left(\derivb{} + \derivb{U}\right)^k \derivb{U}} =
  \sum_{\{\alpha\}} P_\alpha \moyen{\prod_{j=1}^{k+1} \derivmb{U}{\alpha_j}}
\end{equation}
where $\alpha$ is a multi-index and $|\alpha|=k+1$, and $P_\alpha$ a
multiplicity coefficient.
The highest order term of this expression evaluates to
$\sum_{ij} L_i L_j K_{ij} \derivmb{J_{ij}^*}{j} = \derivmb{S}{k}$.

Due to the structure of $U$, spin dependence in (\ref{ss}) will 
come either from the lower derivatives of $J_{ij}^*$ (of the 
form~(\ref{forma}) by hypothesis), from the derivatives of $\lambda_i^*$, or 
explicitly from $U$. In the later case we get a multiplicative 
factor $(\sigma_i-m_i)$. Hence we end up with computing a term, with
$k\ge 1$, of the form
\begin{equation}
  \label{moyes}
\moyen{(\sigma_i-m_i)^k} = (-1)^k (1-m_i^2) \frac{(m+1)^{k-1} - (m-1)^{k-1}}{2}
\end{equation}
Clearly any term including $(\sigma_i-m_i)$ will give a multiplicative
factor $L_i$ after averaging. As spins are decoupled in the $\beta=0$
limit we obtain the product of those factors over the spins in the
diagram as claimed.


\section{Double-counting}
\label{double-counting}

We want to remove two-spin diagrams from the resummation of loop
diagrams.
These two-spin diagrams are precisely the ones appearing in the 
loop diagrams in a system including two spins only. In the case
of $N=2$ spins the matrix $M$ reads 
\begin{equation}
\mathbf{M} = \beta \left(
\begin{array}{cc}
0 & K_{ij}\sqrt{L_i L_j} \\
K_{ij}\sqrt{L_i L_j} & 0
\end{array}
\right) \ .
\end{equation}
We then calculate $J_{ij}^{*(\text{loop})}$ for this simple $N=2$
spin model from formula (\ref{tap}),
and get this way the contribution to be subtracted to the 
sum of 2-spin and loop diagrams (third term in (\ref{duplip})).

\end{document}